\documentclass[aps,prl,amsmath,amssymb,floatfix,twocolumn,superscriptaddress,letterpaper]{revtex4-1}
\usepackage{multirow}
\usepackage{bbold}
\usepackage{subfigure}
\usepackage{color}
\usepackage{mathrsfs}
\usepackage{hyperref}
\usepackage[normalem]{ulem}
\usepackage{bm}

\usepackage[utf8]{inputenc}

\usepackage{amsfonts, relsize, color}
\usepackage{graphics}
\usepackage{graphicx}
\usepackage{subfigure}

\usepackage{ulem}

\def\ket#1{|#1\rangle }
\def\bra#1{\langle #1 |}
\def\braket#1{\langle #1 \rangle}

\def\nn{\nonumber \\ }

\renewcommand{\Im}{\operatorname{Im}}
\newcommand{\ah}{\operatorname{AH}}
\newcommand{\pt}{\operatorname{P}}
\newcommand{\npt}{\operatorname{NP}}
\newcommand{\oo}{\operatorname{O}}
\newcommand{\CP}{\operatorname{CP}}
\newcommand{\LP}{\operatorname{LP}}

\begin{document}
\title{\bf Nonperturbative Topological Current in Weyl and Dirac Semimetals in Laser Fields}

\author{Renato M. A. Dantas}\email{rmad@pks.mpg.de}
\affiliation{Max-Planck-Institut f\"{u}r Physik komplexer Systeme and
	W\"{u}rzburg-Dresden Cluster of Excellence ct.qmat, N\"{o}thnitzer Str. 38,
	01187 Dresden, Germany}

\author{Zhe Wang}
\affiliation{Institute of Physics II, University of Cologne, Cologne, Germany}

\author{Piotr Sur\'owka}\email{surowka@pks.mpg.de}
\affiliation{Max-Planck-Institut f\"{u}r Physik komplexer Systeme and
	W\"{u}rzburg-Dresden Cluster of Excellence ct.qmat, N\"{o}thnitzer Str. 38,
	01187 Dresden, Germany}

\author{Takashi Oka}\email{oka@pks.mpg.de}
\affiliation{The Institute for Solid State Physics, The University of Tokyo, Kashiwa, Chiba 277-8581, Japan}
\affiliation{Max-Planck-Institut f\"{u}r Physik komplexer Systeme, N\"{o}thnitzer Str. 38, 01187 Dresden, Germany}
\affiliation{Max Planck Institute for Chemical Physics of Solids,Dresden 01187, Germany}

\date{\today}
\begin{abstract}
We study non-perturbatively the anomalous Hall current and its high harmonics generated in Weyl and Dirac semimetals by strong elliptically polarized laser fields, in the context of kinetic theory. We find a novel crossover between perturbative and non-perturbative regimes characterized by the electric field strength $\mathcal{E}^{*}= \frac{\mu \omega}{ 2 e v_F}$ ($\omega$: laser frequency, $\mu$: Fermi energy, $v_F$: Fermi velocity). In the perturbative regime, the anomalous Hall current quadratically depends on the field strength ($\mathcal{E}$), whereas the higher order corrections, as well as high harmonics, vanish at zero temperature. In the non-perturbative regime, the anomalous Hall current saturates and decays as $(\log{\mathcal{E}})/\mathcal{E}$, while even-order high harmonics are generated when inplane rotational symmetry is broken. Based on the analytical solution of the Boltzmann equation, we reveal the topological origin of the sharp crossover: the Weyl monopole stays inside or moves outside of the Fermi sphere, respectively, during its fictitious motion in the pertubative or non-pertubative regimes. Our findings establish a new non-linear response intrinsically connected to topology, characteristic to Weyl and Dirac semimetals.
\end{abstract}

\maketitle

\textit{Introduction} - Studies of electronic transport in quantum materials have not only led to major technological advancements in the past decades, but also substantially extended our understanding of novel transport phenomena beyond the conventional Drude paradigm.
A prominent example is the discovery of topological materials, such as Dirac and Weyl semimetals, which are featured by massless chiral quasiparticles at low energies \cite{RevModPhys.90.015001,annurev-matsci-070218-010049}.
Via a microscopic mechanism linked to quantum anomalies \cite{Landsteiner2016}, the topological nature of these quasiparticles can lead to robust transport properties against Ohmic dissipation.
The notion of Berry phases in momentum space plays a central role in the semiclassical description of the topological semimetals \cite{PhysRevLett.95.137204,Duval,RevModPhys.82.1959,Sie2014}. Through the kinetic equation the Berry-phase contributions have been incorporated in the Fermi liquid framework to describe transport phenomena \cite{Loganayagam:2012,PhysRevLett.109.181602,PhysRevLett.109.162001,PhysRevB.88.104412,PhysRevLett.113.182302,RMAD2018}.

Going beyond transport phenomena, the Dirac and Weyl semimetals have recently been found to be a new versatile platform to investigate nonlinear and nonperturbative optical phenomena \cite{Mikhailov_2008,PhysRevB.89.041408,Giorgianni2016,Yoshikawa736,Hafez2018,McIver2019,Oka2019,RMAD2020,PhysRevLett.124.117402}.
Especially at low energies such as in the terahertz (THz) frequency range (1~THz $ \sim 4$~meV), highly efficient odd-order high-harmonic generation (HHG) was very recently reported in graphene \cite{PhysRevB.89.041408,Hafez2018} and the three-dimensional Dirac semimetal Cd$_3$As$_2$ \cite{RMAD2020,PhysRevLett.124.117402}.
In particular, the observed nonperturbative HHG induced by linearly polarized light (LPL) has been successfully described by kinetic theory, which established the driven kinetics of the Dirac fermions as the microscopic mechanism for the nonperturbative HHG in Cd$_3$As$_2$ \cite{RMAD2020}.

%%%%%%%%%%%%%%%%%%%%%%%%%%%%%
%%%%%%%%%%%%%%%%%%%%%%%%%%%%%
\begin{figure}[t!]
	\includegraphics[width=0.7\linewidth]{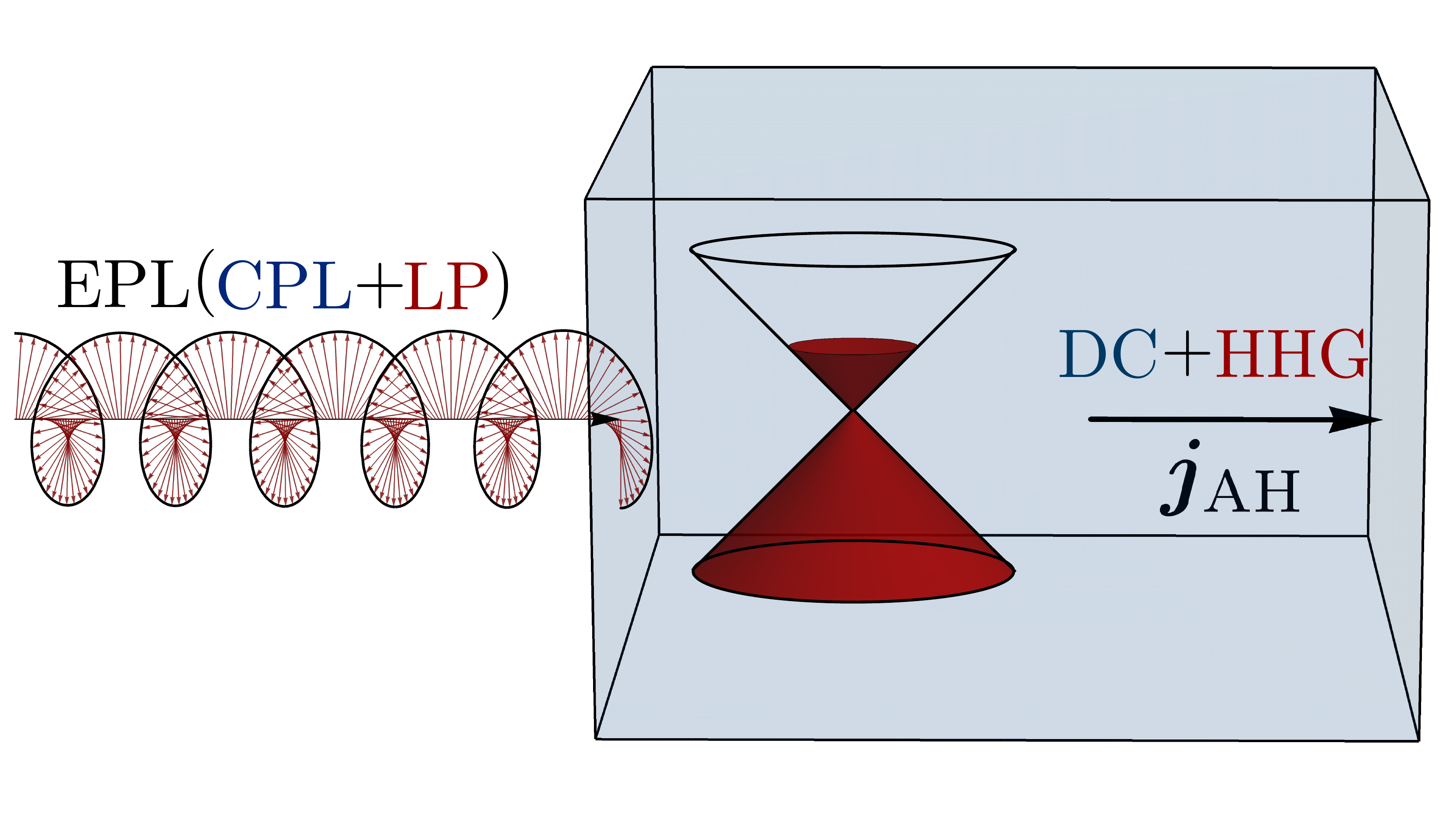}
	\caption{Anomalous Hall current ($\bm j _{\ah}$) in a Weyl semimetal (WSM) induced by elliptically polarized light (EPL). The circularly polarized component (CPL) of EPL generates a DC anomalous Hall current, while an additional linearly polarized (LP) component induces even order topological high harmonic generation (HHG).}
	\label{fig:setup}
\end{figure}
%%%%%%%%%%%%%%%%%%%%%%%%%%%%%
%%%%%%%%%%%%%%%%%%%%%%%%%%%%%

In this work, motivated by these recent experimental advances \cite{RMAD2020,PhysRevLett.124.117402}, we perform kinetic theory analysis of the topological nonlinear response of the Weyl semimetals (WSM) driven by circularly (CPL) and elliptically polarized laser light (EPL) with experimentally relevant model parameters. 
As depicted in Fig.~\ref{fig:setup}, we show that the laser field generates an anomalous Hall current
flowing in the direction of the laser propagation. 
Our analysis is extended from the perturbative regime to the strong field nonperturbative regime. 
We obtain an explicit analytical expression of the zero-temperature anomalous Hall current, and find  
even-order HHG when Weyl's SO(3) symmetry is broken, as in the case of EPL. 
The mechanism of the even-order HHG is topological and nonperturbative, which 
we reveal by identifying the dependence of the anomalous Hall current on the Berry curvature. 
Further numerical analysis indicates that the zero-temperature analytical results can be valid even for low but finite temperatures.
We note that our analysis also applies to Dirac semimetals which consist of two degenerate Weyl fermions with Berry curvatures that have opposite signs, and the anomalous Hall current becomes a relative flow, e.g. spin current.

\textit{General Theory} - Our analysis employs kinetic theory within the relaxation time approximation. This approach is reliable when quantum transitions are irrelevant, which is the case when a system with a finite Fermi sphere is driven by THz lasers \cite{RMAD2020}. In this framework, the Boltzmann kinetic equation 
%%%%%%%%%%%%%%%%%%%%%%%%%%%%%
%%%%%%%%%%%%%%%%%%%%%%%%%%%%%
\begin{equation}
\partial_t f+\dot{\bm{r}} \cdot\nabla_{\bm{r}} f + \dot{\bm{p}} \cdot \nabla_{\bm{p}} f  =\tau^{-1} (f_0 -f), 
\label{eq:BE1}
\end{equation}
%%%%%%%%%%%%%%%%%%%%%%%%%%%%%
%%%%%%%%%%%%%%%%%%%%%%%%%%%%%
describes the evolution of the distribution function $f$ in phase space. $f_0$ stands for the equilibrium Fermi-Dirac distribution function and $\tau$ for the relaxation time. The effective dynamics of the Weyl quasiparticles is determined by the modified equations of motion \cite{RevModPhys.82.1959,PhysRevB.88.104412,PhysRevLett.109.162001,PhysRevB.94.245121}
%%%%%%%%%%%%%%%%%%%%%%%%%%%%%
%%%%%%%%%%%%%%%%%%%%%%%%%%%%%
\begin{equation}
\dot{\bm{r}} = \nabla_{\bm{p}} \epsilon_{\bm{p}}-\hbar \dot{\bm{p}} \times \bm{\Omega}_{\bm{p}},
\qquad \dot{\bm{p}}= -e\bm{E} - e \dot{\bm{r}}  \times \bm{B},
\label{eq:EOM1}
\end{equation}
 %%%%%%%%%%%%%%%%%%%%%%%%%%%%%
 %%%%%%%%%%%%%%%%%%%%%%%%%%%%%
 where $ \mathbf{\Omega}_{\mathbf{p}}= - \Im[\bra{\nabla_{\bm p} u_{\bm p}} \times \ket{\nabla_{\bm p} u_{\bm p}}] $ represents the Berry curvature, $\epsilon_{\mathbf{p}} $ the energy dispersion relation, and $\bm{E} $ and $\bm{B}$ the electric and magnetic field, respectively.

Hereafter, we focus on homogeneous responses and assume no external magnetic field. Additionally, we neglect the magnetic field induced by the dominant oscillatory electric field driving the system. These considerations lead to the simplified Boltzmann equation 
%%%%%%%%%%%%%%%%%%%%%%%%%%%%%
%%%%%%%%%%%%%%%%%%%%%%%%%%%%%
\begin{equation}
\left(\tau \partial_t  + 1\right)f- \tau e \bm E \cdot \nabla_{\bm p} f =f_0 ,
\label{eq:BE2}
\end{equation}
%%%%%%%%%%%%%%%%%%%%%%%%%%%%%
%%%%%%%%%%%%%%%%%%%%%%%%%%%%%
and current density
%%%%%%%%%%%%%%%%%%%%%%%%%%%%%
%%%%%%%%%%%%%%%%%%%%%%%%%%%%%
\begin{equation}
\bm {j}=-e\int_{p}  \nabla_{\bm{p}} \epsilon_{\bm{p}} \, f - e^2 \hbar \,  \bm E \times \int_{p} \bm \Omega_{\bm p} f \\
\label{eq:C2}
\end{equation}
%%%%%%%%%%%%%%%%%%%%%%%%%%%%%
%%%%%%%%%%%%%%%%%%%%%%%%%%%%%
where $\int_{p} \equiv \int\frac{d^3 p}{(2\pi \hbar)^3}$. The current density can be decomposed as $\bm {j}=\bm {j}_{\oo}+\bm {j}_{\ah}$, where $\bm j _{\oo}$ corresponds to the usual current generated by the group velocity (first term in Eq.~(\ref{eq:C2})), while $\bm j _{\ah}$ (second term) corresponds to the anomalous Hall current. The anomalous Hall current is topological as it is induced by the expectation value of the Berry curvature.

The crucial step towards computing the response current is to solve the Boltzmann equation [Eq.~(\ref{eq:BE2})] with the boundary condition $f(t=0,\mathbf{p})=f_0(\mathbf{p})$. The solution can be shown, either by the method of characteristics or Fourier transform, to take the following form
%%%%%%%%%%%%%%%%%%%%%%%%%%%%%
%%%%%%%%%%%%%%%%%%%%%%%%%%%%%
\begin{eqnarray}
f(t,\bm{p})&=& e^{-t/\tau} f_0(\bm{p}-e\bm{A}(t)) \nn
&&+\frac{1}{\tau}\int_0^t ds\, e^{\frac{s-t}{\tau}} f_0(\bm p -e \bm \Delta (t,s)),
\label{eq:StBE}
\end{eqnarray}
%%%%%%%%%%%%%%%%%%%%%%%%%%%%%
%%%%%%%%%%%%%%%%%%%%%%%%%%%%%
where $\bm A (t)=-\int^t_0 ds\, \bm E (s)$ is the vector potential and $\bm \Delta (t,s) = \bm A (t)-\bm A (s)$.  The first term of this equation corresponds to the Fermi-Dirac distribution shifted by the vector potential $\bm A (t)$. Since this term vanishes exponentially fast with $t/\tau$, it is only relevant for large values of the relaxation time $\tau$. In fact, $f(t,\bm{p})=  f_0(\bm{p}-e\bm{A}(t)) $ is the solution of the collision-less ($\tau \rightarrow \infty$) Boltzmann equation. The second term of Eq.~(\ref{eq:StBE}) incorporates collisions in our description. This term averages over the Fermi-Dirac distributions shifted by $\bm \Delta (t,s)$,
and we can portrayed it as resulting from a fictitious motion of the Fermi sphere centered at $\bm \Delta (t,s)$, 
where the contribution is exponentially suppressed with $(t-s)/\tau$. In this construction, $\tau$ effectively defines how much the driving electric field can deform the equilibrium distribution.

So far, the derived formalism is not system specific. Henceforth, we focus on the anomalous Hall current generated due to a single Weyl node (Fig.~\ref{fig:setup}) with the Hamiltonian $ H=\eta \, v_F \bm \sigma \cdot \bm p$, where $\eta=\pm 1$ defines the chirality and $v_F$ is the Fermi velocity. The Pauli matrices $\bm \sigma=(\sigma_x,\sigma_y,\sigma_z)$ act on the pseudo-spin indices and momentum $\bm p$ is measured from the Weyl node. The energy dispersion relation and Berry curvature for the conduction band are given, respectively, by $ \epsilon_{\bm p} =  v_F p$ and $\bm \Omega _{\bm p}= - \eta \hat{\bm p}/2 p^2$. We denote the Fermi energy by $\mu$.

\textit{Circularly Polarized Light} -  We investigate the anomalous Hall response current induced by CPL propagating in the $\bm{\hat{z}}$-direction, i.e. $\bm E (t)= \mathcal{E} \cos (\omega t) \bm{\hat{x}}+\mathcal{E} \sin (\omega t) \bm{\hat{y}}$. 
Exploiting the fact that CPL respects Weyl's spherical symmetry, we first show that the resulting $\bm j _{\ah}$ is constant in time. Later, we compute the analytical form of $\bm j _{\ah}$ and explain its perturbative to non-perturbative crossover in terms of the fictitious motion of the Weyl monopole.

%%%%%%%%%%%%%%%%%%%%%%%%%%%%%
%%%%%%%%%%%%%%%%%%%%%%%%%%%%%
\begin{figure}[t!]
	\includegraphics[width=0.95\linewidth]{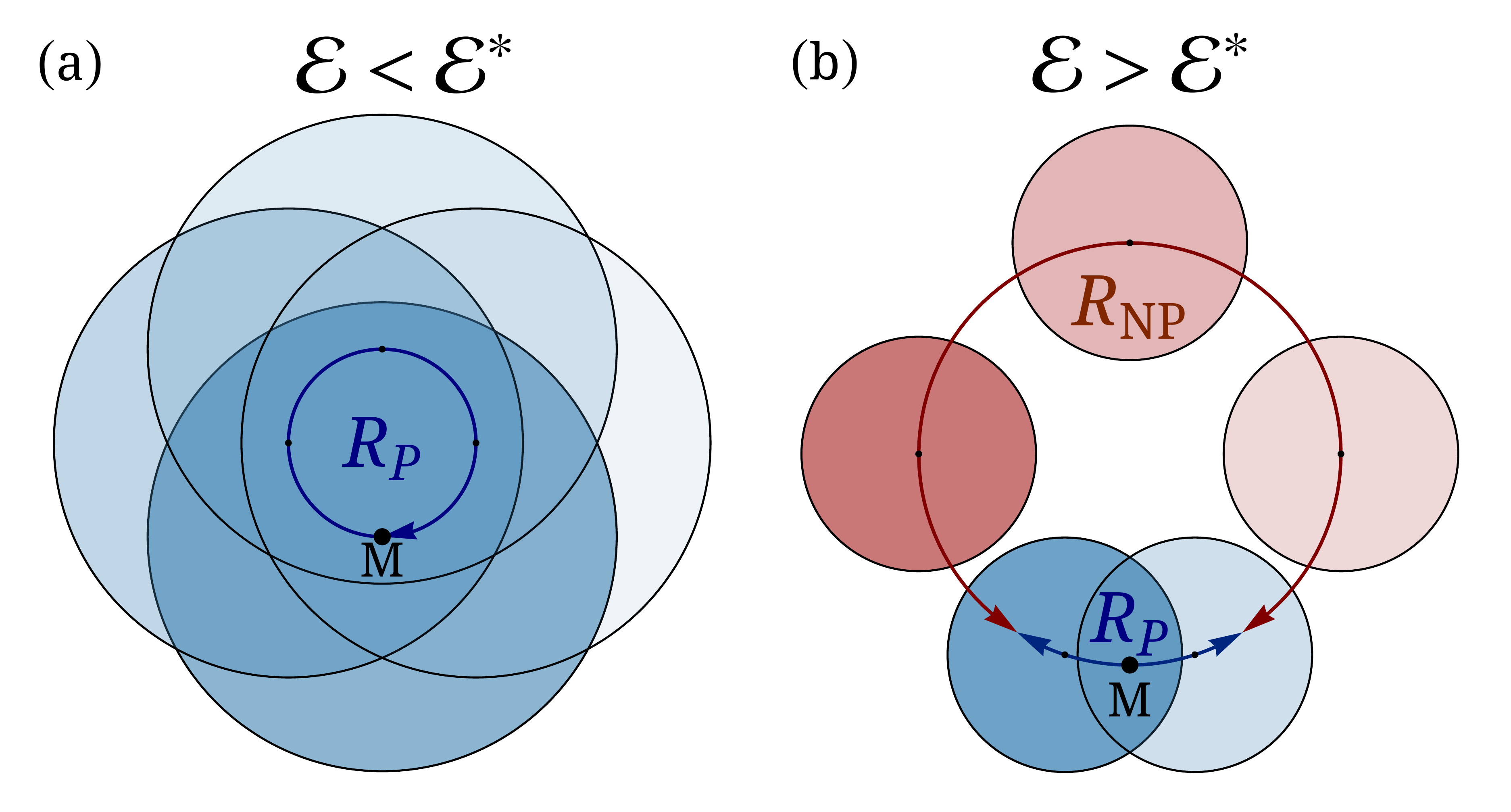}
	\caption{Illustration of the shifted distribution functions $f_0(\bm p -e \bm \Delta(0,u))$ contributing to the anomalous Hall current, $\bm j _{\ah}$ [Eq.~(\ref{eq:C_CP3})], induced by CPL in the (a) weak ($\mathcal{E}<\mathcal{E}^*$) and (b) strong ($\mathcal{E}>\mathcal{E}^*$) field regimes. $\mathcal{E}^*$ is maximum field value for which the monopole remains enclosed by the Fermi surface during its fictitious motion [Eq.~(\ref{eq:es})]. Circles represent the Fermi sphere shifted by $\bm \Delta(0,u)= \bm A(0)- \bm A(u)$. In the weak field regime, the overlap between shifted distribution and Weyl monopole M enables perturbation theory to render the exact result of the current. In the strong field regime, perturbation theory is only able to describe a small set of shifted distributions, denoted by $R_{\pt}$ (depicted by the blue arrow). The remaining distributions contributing to the current belong to $R_{\npt}$ (red arrow) and are responsible for the non-perturbative behaviour of $\bm j_{\ah}$. }
	\label{fig:P2NP}
\end{figure}
%%%%%%%%%%%%%%%%%%%%%%%%%%%%%
%%%%%%%%%%%%%%%%%%%%%%%%%%%%%

For simplicity, we consider a system that has been driven long enough such that it has effectively lost information of the initial state. This consideration, along with the $T=2 \pi/\omega$ periodicity imposed by the driving electric field, allows us to write the anomalous Hall current as  
%%%%%%%%%%%%%%%%%%%%%%%%%%%%%
%%%%%%%%%%%%%%%%%%%%%%%%%%%%%
\begin{equation}
\bm j _{\ah}= \frac{- e^2 \hbar}{1-e^{-T/\tau}} \int_{-T}^{0} \frac{du}{\tau} \, e^{\frac{u}{\tau}} \bm E(t) \times \braket{\bm \Omega (t,u+t)} ,
\label{eq:C_CP}
\end{equation}
%%%%%%%%%%%%%%%%%%%%%%%%%%%%%
%%%%%%%%%%%%%%%%%%%%%%%%%%%%%
with $\braket{\bm \Omega (t,u+t)} = \int_{p} \bm \Omega_{\bm p}  f_0(\bm p -e\bm \Delta (t,u+t))$. The anomalous Hall current $\bm j _{\ah}$ points in the $\bm{\hat{z}}$ direction since $\bm E(t)$ and  $\braket{\bm \Omega (t,u+t)}$ lie within the $p_x-p_y$ plane. Thus, it is useful to move to the frame co-rotating with the electric field ($\bm E (t)= \bm R (t) \bm E$, where $\bm R (t)$ is the rotation matrix in the $x-y$ plane and $\bm E= \mathcal{E} \bm{\hat{x}}$) and use the rotational invariance of $\epsilon_{\bm p}$, to obtain 
%%%%%%%%%%%%%%%%%%%%%%%%%%%%%
%%%%%%%%%%%%%%%%%%%%%%%%%%%%%
\begin{equation}
	\bm j _{\ah}=- \frac{e^2 \hbar}{\tau} \frac{1}{1-e^{-T/\tau}} \int_{-T}^{0} du \, e^{\frac{u}{\tau}}  \bm E \times \braket{\bm \Omega (0,u)},
	\label{eq:C_CP3}
\end{equation}
%%%%%%%%%%%%%%%%%%%%%%%%%%%%%
%%%%%%%%%%%%%%%%%%%%%%%%%%%%%
which is time independent. Thereby, CPL can only induce a DC anomalous Hall current. 

The anomalous Hall current can be computed resorting to the zero temperature expectation value of the Berry curvature 
%%%%%%%%%%%%%%%%%%%%%%%%%%%%%
%%%%%%%%%%%%%%%%%%%%%%%%%%%%%
\begin{equation}
	\braket{\bm \Omega (0,u)} = 
	\begin{cases}
		- \frac{\eta}{12 \pi^2 \hbar^3} e \bm \Delta (0,u) , &   \| e \bm \Delta (0,u) \| \le \mu/v_F \\
		\\
		\frac{1}{6 \pi^2 \hbar^3}(\frac{\mu}{v_F})^3 \bm \Omega_{e \bm \Delta (0,u)},& \| e \bm \Delta (0,u) \|> \mu/v_F \\
	\end{cases},
	\label{eq:C_CP5}
\end{equation}
%%%%%%%%%%%%%%%%%%%%%%%%%%%%%
%%%%%%%%%%%%%%%%%%%%%%%%%%%%%
which can be understood in terms of the relative position of the Weyl monopole to the Fermi surface. If the monopole is enclosed by the Fermi surface, $\braket{\bm \Omega (0,u)} $ derives from the linear term in perturbation theory, whereas higher order corrections cancel out. On the other hand, when the monopole is outside the Fermi surface, perturbation theory breaks down as $\braket{\bm \Omega (0,u)}$ becomes non-perturbative. This result leads to a topological crossover between perturbative and non-perturbative regimes in $\bm j _{\ah}$, characterized by the critical field strength
%%%%%%%%%%%%%%%%%%%%%%%%%%%%%
%%%%%%%%%%%%%%%%%%%%%%%%%%%%%
\begin{equation}\label{eq:es}
	\mathcal{E}^{*}= \frac{\mu \omega}{ 2 e v_F}.
\end{equation}  
%%%%%%%%%%%%%%%%%%%%%%%%%%%%%
%%%%%%%%%%%%%%%%%%%%%%%%%%%%%
For reference, we note that the critical field takes value of $\mathcal{E}^{*}= 4.75\mbox{kV/cm}$ for the material parameters $\mu=118\mbox{meV},\;v_F=7.8\times 10^5\mbox{m/s},\;\omega/2\pi =4\mbox{meV}$ (1THz) used in Ref.\cite{RMAD2020}, which is an experimentally accessible field strength. 

In the weak field regime ($\mathcal{E}<\mathcal{E}^*$), the Weyl monopole is always enclosed by the Fermi surface during its fictitious motion [Fig.~\ref{fig:P2NP}(a)] and the evaluation of Eq.~(\ref{eq:C_CP3}) entails 
%%%%%%%%%%%%%%%%%%%%%%%%%%%%%
%%%%%%%%%%%%%%%%%%%%%%%%%%%%%
\begin{equation}
\bm {j}_{\ah} =\eta \frac{e^3}{3(2\pi \hbar)^2}  \frac{\omega \tau^2}{1+(\omega \tau)^2} \mathcal{E}^2 \bm{\hat{z}},
\label{eq:IF}
\end{equation}
%%%%%%%%%%%%%%%%%%%%%%%%%%%%%
%%%%%%%%%%%%%%%%%%%%%%%%%%%%%
which is equivalent to the result achieved by solving the Boltzmann equation [Eq.~(\ref{eq:BE2})] perturbatively in the electric field \cite{PhysRevLett.115.216806}. Nevertheless, it should be stressed that there are no higher order corrections, so the second order result of Eq.~(\ref{eq:IF}) is exact.

In the strong field regime ($\mathcal{E}>\mathcal{E}^*$), the Weyl monopole moves out of the Fermi surface during its fictitious motion defining two physically distinct regimes when performing the integral in $u$ [Eq.~(\ref{eq:C_CP3})]: the perturbative ($R_{\pt}$) and the non-perturbative regime ($R_{\npt}$) (Fig.~\ref{fig:P2NP}(b)). Eq.~(\ref{eq:C_CP3}) can then be analytically evaluated, leading to the zero temperature non-perturbative result
%%%%%%%%%%%%%%%%%%%%%%%%%%%%%
%%%%%%%%%%%%%%%%%%%%%%%%%%%%%
\begin{equation}
\bm j _{\ah} = \eta \frac{e^2 }{(2 \pi \hbar)^2} \frac{ \bm{\hat{z}} }{1-e^{-\frac{T}{\tau}}}\left[ \xi_{\pt}\bigg|^{0}_{-u_1^*} +\xi_{\pt}\bigg|^{-u_2^*}_{-T} + \xi_{\npt}\bigg|_{-u_2^*}^{-u_1^*}\right] 
\label{eq:C_CP_A}
\end{equation}
%%%%%%%%%%%%%%%%%%%%%%%%%%%%%
%%%%%%%%%%%%%%%%%%%%%%%%%%%%%
where
%%%%%%%%%%%%%%%%%%%%%%%%%%%%%
%%%%%%%%%%%%%%%%%%%%%%%%%%%%%
\begin{eqnarray}~\label{eq:xi_p}
\xi_{\pt}&=& \frac{ e \mathcal{E}^2 e^{\frac{u}{\tau} } \left[1 +\tau ^2 \omega ^2-\cos (u \omega ) -\tau  \omega \sin (u \omega )\right]}{3\omega  \left(1+\tau ^2 \omega ^2 \right)}, \\
\xi_{\npt}&=&  \frac{ \mu ^3 \omega ^2 e^{ \frac{u}{\tau }+\frac{i u \omega }{2}} \, _2F_1\left[1,\frac{1}{2}-\frac{i}{\tau  \omega };\frac{3}{2}-\frac{i}{\tau  \omega };e^{i u \omega }\right]}{3 e^2 v_F^3 \mathcal{E} (\tau  \omega -2 i)} , ~\label{eq:xi_np}
\end{eqnarray}
%%%%%%%%%%%%%%%%%%%%%%%%%%%%%
%%%%%%%%%%%%%%%%%%%%%%%%%%%%%
$u_1^*= \omega^{-1} \arccos \left(1-\frac{\mu ^2 \omega ^2}{2 e^2 v_F^2 \mathcal{E}^2}\right)$, $u_2^*=T-u_1^*$ and $_2F_1$ is the Gaussian hypergeometric function. It should be noted that in the strong field regime ($\mathcal{E}>\mathcal{E}^*$), even the terms arising from perturbation theory ($\xi_{\pt}$) acquire non-perturbative character due to the dependence of $u^*_1$ and $u^*_2$ on the electric field.
%%%%%%%%%%%%%%%%%%%%%%%%%%%
%%%%%%%%%%%%%%%%%%%%%%%%%%%
\begin{figure}[t!]
	\includegraphics[width=0.95\linewidth]{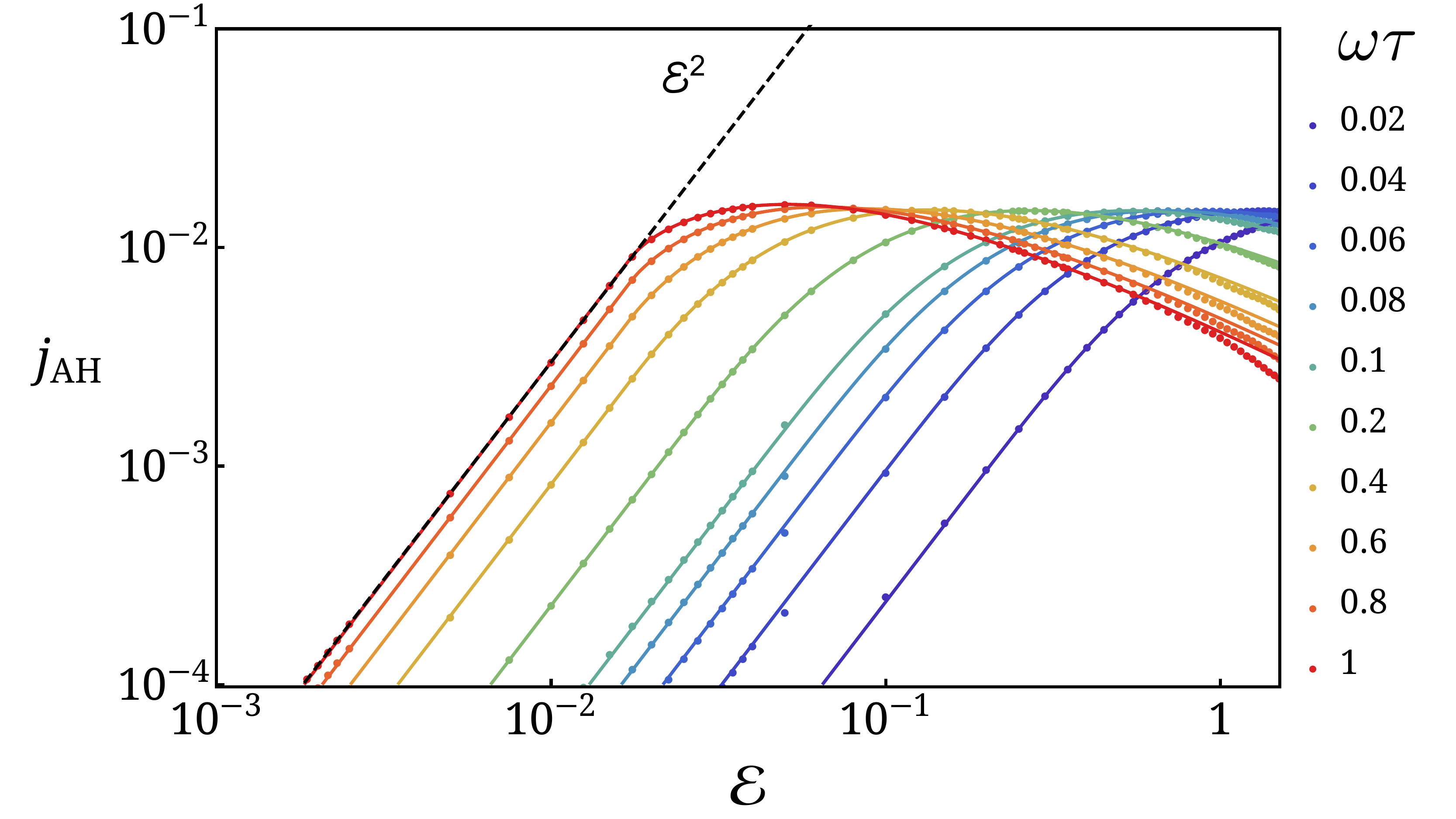}
	
	\caption{ Anomalous Hall current density ($\bm j _{\ah}$) for a single Weyl node ($\eta=1$) at 30 K, measured in units of $e \mu^3/ v^2_F (2 \pi \hbar)^3$, as function of electric field $\mathcal{E}$, in units of $\mu^2/ e  \hbar v_F $, for different values of $\omega \tau$. Circles result from numerical evaluation of Eq.~(\ref{eq:C_CP}), while solid lines stem from the analytical results presented in Eqs.~(\ref{eq:IF}),(\ref{eq:C_CP_A})-(\ref{eq:xi_np}). For numerical analysis, we set $\mu =118\,\mathrm{meV}$ and $v_F=7.8\times 10^5\mbox{m/s}$.
	}~\label{Fig:CPL}
\end{figure}
%%%%%%%%%%%%%%%%%%%%%%%%%%%
%%%%%%%%%%%%%%%%%%%%%%%%%%%
%%%%%%%%%%%%%%%%%%%%%%%%%%%%%%%%%%%%%%%%%%%%%%%%%%%%%%%%%%%%%%%%%%%%%%%
\begin{figure*}[t!]
	\includegraphics[width=0.99\linewidth]{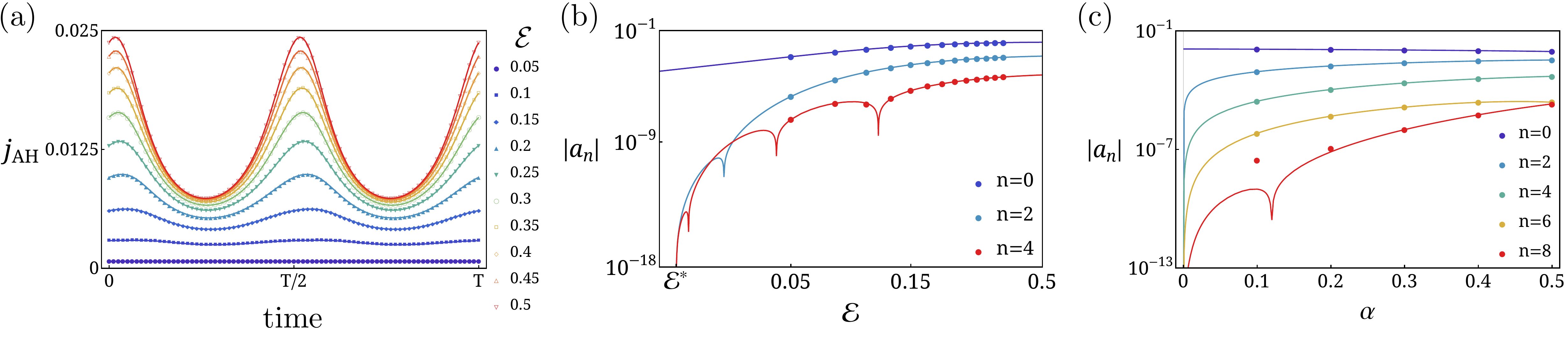}
	\caption{ (a) Time dependence of the anomalous Hall current density ($\bm j _{\ah}$) for a single Weyl node ($\eta=1$) at $30$ K, measured in units of $e \mu^3/ v^2_F (2 \pi \hbar)^3$, for a period T of the driving EPL with $2 \mathcal{E}_x =\mathcal{E}_y=\mathcal{E}$ ($\alpha=0.5$), for different values of $\mathcal{E}$ (in units of $\mu^2/ e  \hbar v_F$) and $\omega \tau=0.1$. (b) Field dependence of HHG, defined as an absolute value cosine Fourier coefficient $|a_n|$ of $\bm j_{\ah}$, for $n=0,2$ and $4$ with $\alpha=0.1$ at $30$ K. Circles result from numerical evaluation of Eq.~(\ref{eq:C_CP}) for EPL, while solid lines are obtained from analytical evaluation of $\bm j _{\ah}$ up to cubic order in $\alpha$. (c) Ellipticity dependence of HHG with $\mathcal{E}=0.25$ at $30$ K. Circles result from numerical evaluation of Eq.~(\ref{eq:C_CP}), while solid lines are obtained from analytical evaluation of $\bm j _{\ah}$ up to tenth order in $\alpha$. For numerical analysis, we considered $\mu =118\,\mathrm{meV}$ and $v_F=7.8\times 10^5\mbox{m/s}$.
	}~\label{Fig:EPL}
\end{figure*}
%%%%%%%%%%%%%%%%%%%%%%%%%%%%%%%%%%%%%%%%%%%%%%%%%%%%%%%%%%%%%%%%%%%%%%%
Despite the fact that our analytical results have been derived assuming zero temperature, numerical evaluation of Eq.~(\ref{eq:C_CP}) reveals excellent agreement even at 30 K (Fig.~\ref{Fig:CPL}). These results attest the validity of Eqs.~(\ref{eq:IF}) and (\ref{eq:C_CP_A}) for low finite temperatures.

\textit{High-Harmonic Generation} - We now look into the anomalous Hall current induced by EPL propagating in the $\bm{\hat{z}}$-direction. In contrast to CPL, we show that Weyl's SO(3) symmetry breaking induced by EPL leads to high harmonics generation in $\bm j _{\ah}$. The HHG is suppressed for weak fields and only occurs in the $\mathcal{E}>\mathcal{E}^{*}$ regime. Furthermore, we present the analytical treatment for $\bm j _{\ah}$ and demonstrate that the generated harmonics must be even and non-perturbative in $\mathcal{E}$. 

With the intention of performing perturbation theory around CPL, we consider the field $\bm E(t)=\mathcal{E} (1-\alpha) \cos (\omega t) \bm{\hat{x}}+\mathcal{E} \sin (\omega t) \bm{\hat{y}}$, where $\alpha$ plays the role of a symmetry breaking parameter. In this way, we decompose electric field and vector potential in their circularly (CP) and linearly (LP) polarized components. Hence, similarly to CPL, we use Weyl's SO(3) symmetry to rewrite Eq.~(\ref{eq:C_CP}) as
%%%%%%%%%%%%%%%%%%%%%%%%%%%%%
%%%%%%%%%%%%%%%%%%%%%%%%%%%%%
\begin{eqnarray}
	&&\bm j _{\ah}=\eta \frac{e^2 \hbar}{\tau} \frac{1}{1-e^{-T/\tau}} \int_{-T}^0 du e^{\frac{u}{\tau}} \int_{p}  f_0(\bm p -e \bm \Delta_{\CP} (0,u)) \nn
	&&\frac{ (\bm E _{\CP}(t) + \alpha \bm E _{\LP} (t) ) \times (\bm R (t)\bm p +\alpha e \bm \Delta_{\LP} (t,u+t))}{2 | \bm R (t) \bm p + \alpha e  \bm \Delta_{\LP} (t,u+t)| ^3},
	\label{eq:C_EP2}
\end{eqnarray}
%%%%%%%%%%%%%%%%%%%%%%%%%%%%%
%%%%%%%%%%%%%%%%%%%%%%%%%%%%%
which can then be expanded in $\alpha$. While the zero order term corresponds to the result obtained for CPL [Eq.~(\ref{eq:C_CP3})], the linear term contains second order harmonics. Moreover, one can verify that higher order terms in the expansion contain higher order harmonics. Nonetheless, due to the nature of the expansion, those harmonics are always even. This is in sharp contrast to HHG in $\bm {j}_{\oo}$ which is dominated by odd harmonics. This observation is consistent with the $T/2$ periodicity of $\bm j_{\ah}$ [Fig.~\ref{Fig:EPL}(a)] and sets forth a distinctive feature of topological HHG stemming from the fact that $\bm j_{\ah}$, contrary to $\bm {j}_{\oo}$, has an explicit dependence on $\bm E (t)$ [Eq.~(\ref{eq:C2})].

Numerical evaluation of $\bm j_{\ah}$ induced by EPL for finite temperatures supports the conclusions derived from symmetry considerations (Fig.~\ref{Fig:EPL}). The results reveal an intense generation of non-perturbative even order topological high harmonics for $\mathcal{E}>\mathcal{E}^*$ [Fig.~\ref{Fig:EPL}(b)], which grows with the ellipticity ($\alpha$) of the laser fields [Fig.~\ref{Fig:EPL}(c)].

Up to this point, we resorted to symmetry arguments to disclose the high harmonic structure contained in the anomalous Hall response of a Weyl node driven by EPL. The specific form of the current in the strong field regime ($\mathcal{E}>\mathcal{E}^*$) can be computed as a power series in $\alpha$ using the analytical treatment presented for CPL [Eqs.~(\ref{eq:C_CP5})-(\ref{eq:C_CP_A})]. In this approach Eq.~(\ref{eq:C_CP5}) remains valid under the substitutions $\braket{\bm \Omega (0,u)} \rightarrow \braket{\bm \Omega (t,u+t)}$, $\bm \Delta(0,u) \rightarrow \bm \Delta(t,u+t)$. Eq.~(\ref{eq:C_CP_A}) stays unchanged, with $u_1^{*}$ and $u_2^{*}$  representing, respectively,  the smallest and largest solutions of $v_F \|e \bm  \Delta(t,u+t)\|=\mu$, for $u \in [0, T]$~\footnote{Eq.~(\ref{eq:C_CP_A}) trivially generalizes for the existence of multiple crossings between the perturbative and non-perturbative regions, i.e., for the existence of multiple pairs of $u^*_1$ and $u^*_2$.}. $\xi_{\pt}$ and $\xi_{\npt}$ are then computed perturbatively in $\alpha$. The zero order current for EPL is given by Eqs.~(\ref{eq:xi_p})-(\ref{eq:xi_np}), while the linear order reads
%%%%%%%%%%%%%%%%%%%%%%%%%%%%%
%%%%%%%%%%%%%%%%%%%%%%%%%%%%%
\begin{eqnarray}~\label{eq:xi_alpha1_p}
&&\xi^{(1)}_{\npt}=\frac{-  \sqrt{2} \mu ^3 \omega ^2 \sin ^3\left(\frac{u \omega }{2}\right) e^{\frac{u}{\tau}+\frac{i \omega u }{2}}}{e^2 v_F^3 \mathcal{E}  \left(\tau ^2 \omega ^2+4\right) (1-\cos (u \omega ))^{3/2}}  \Big\{  (\tau  \omega +2 i)  \nn
&& \left( \cos 2  \omega t +\frac{1}{3}\right)  \, _2F_1\left[1,\frac{1}{2}-\frac{i}{\tau  \omega };\frac{3}{2}-\frac{i}{\tau  \omega };e^{i u \omega }\right] \nn
&& - e^{- \frac{i \omega u }{2}} \left[\tau  \omega  \cos \left( 2 \omega t+ \frac{u \omega}{2}\right)-2 \sin \left(2 \omega t+ \frac{u \omega}{2}\right)\right] \Big \}, \;\;\;\;\;
~\label{eq:xi_alpha1_np}
\end{eqnarray}
%%%%%%%%%%%%%%%%%%%%%%%%%%%%%
%%%%%%%%%%%%%%%%%%%%%%%%%%%%%
and $\xi^{(1)}_{\pt}=- \xi^{(0)}_{\pt}$. Note that $\xi^{(1)}_{\npt}$ contains the anticipated second harmonic. Furthermore, this method allows the analytical determination of higher order terms in $\alpha$, which contain higher harmonics. It is important to note that, in the presented scheme, there is an implicit dependence on $\alpha$ and $t$ that is introduced by $u^*_{1}$ and $u^*_{2}$. This is the reason why, even though $\xi_{\pt}$ contains only up to linear terms in $\alpha$ and no explicit high harmonics, it also displays HHG. Moreover, $u^*_{1}$ and $u^*_{2}$ also depend on $\mathcal{E}$, ensuring that the HHG associated with both $\xi_{\pt}$ and $\xi_{\npt}$  is non-perturbative in the field strength [Fig.~\ref{Fig:EPL}(b)]. These results should be contrasted with the absence of HHG in the weak field regime ($\mathcal{E}<\mathcal{E}^{*}$), where perturbation theory alone yields Eq.~(\ref{eq:IF}) with an additional multiplicative factor of $(1-\alpha)$. Finally, our analytical treatment presents excellent agreement with the results from numerical evaluation of $\bm j_{\ah}$, even for finite temperatures (Fig.~\ref{Fig:EPL}).

So far, we have limited our discussion to the anomalous Hall response of a single Weyl node. However, realistic WSM must contain multiple Weyl nodes such that, for every Weyl node with chirality $\eta$, exists an anti-Weyl node with opposite chirality ($- \eta$) \cite{NIELSEN1981219,NIELSEN198120,NIELSEN1981173}. In such systems, the total electric current is defined by the sum of the currents generated by each node. On the other hand, the sum of the currents of each node multiplied by its chirality gives the axial or spin current. For this reason, the total electric current associated with the anomalous Hall transport is, in general, finite only for noncentrosymmetric WSM with time reversal symmetry \cite{PhysRevLett.117.216601}. The axial current can be finite for both Dirac and Weyl semimetals.

\textit{Conclusion} - In this work we have discussed the anomalous Hall current in Weyl and Dirac semimetals induced by elliptically polarized laser fields, in the context of kinetic theory. We have identified a topological crossover between perturbative and non-perturbative regimes, characterized by $\mathcal{E}^*$: for $\mathcal{E}<\mathcal{E}^*$, the Weyl monopole stays inside the Fermi surface during its fictitious motion while, for $\mathcal{E}>\mathcal{E}^*$, it moves outside. In the perturbative regime ($\mathcal{E}<\mathcal{E}^*$) the current scales as $\mathcal{E}^2$ and is constant in time for both CPL and EPL. For $\mathcal{E}>\mathcal{E}^*$, the current becomes non-perturbative and, when the Weyl SO(3) symmetry is broken by the laser fields, non-perturbative even order high harmonics connected to the topology of the material are generated. This mechanism generalizes for more complex Weyl Hamiltonians, possibly with anisotropic Fermi velocities, tilted cones and higher charge monopoles \cite{avetissian2020,Huang1180,PhysRevResearch.2.013007}. These systems lack, in general, SO(3) symmetry and even CPL shall induce topological HHG. Numerical analysis for realistic material parameters and field strengths supports our analytical findings, opening the possibility for experimental observation of topological HHG, a novel non-linear response characteristic of Dirac and Weyl semimetals. 

\begin{acknowledgments}
\textit{Acknowledgements} - R.M.A.D. acknowledges useful conversations with Inti Sodemann. R.M.A.D. and P.S. acknowledge financial support by the Deutsche Forschungsgemeinschaft (DFG, German Research Foundation) under Germany's Excellence Strategy through W\"{u}rzburg-Dresden Cluster of Excellence on Complexity and Topology in Quantum Matter - ct.qmat (EXC 2147, project-id 390858490). Z.W. acknowledges support by the DFG via the project No. 277146847—Collaborative Research Center 1238: Control and Dynamics of Quantum Materials (Subproject No. B05). T.O. acknowledges support by JST CREST Grant No. JPMJCR19T3, Japan.
\end{acknowledgments}

\bibliography{THHG}
	
\end{document}